\documentclass[longauth]{aa}
\usepackage{graphicx}
\usepackage{txfonts}
\usepackage{xcolor}
\usepackage{hyperref}
\hypersetup{colorlinks, linkcolor=blue, citecolor=blue, urlcolor=blue}

\newcommand{\muvec}{\mbox{\boldmath $\mu$}}

\newcommand{\te}{t_{\rm E}}
\newcommand{\thetae}{\theta_{\rm E}}

\newcommand{\pie}{\pi_{\rm E}}
\newcommand{\pien}{\pi_{{\rm E},N}}
\newcommand{\piee}{\pi_{{\rm E},E}}
\newcommand{\dl}{D_{\rm L}}
\newcommand{\ds}{D_{\rm S}}

\definecolor{brown}{rgb}{0.59, 0.29, 0.0}
\definecolor{darkgreen}{rgb}{0.0, 0.42, 0.24}
\definecolor{darkblue}{rgb}{0.01, 0.31, 0.59}
\definecolor{darkblue}{rgb}{0.0, 0.25, 0.42}
\definecolor{blue}{rgb}{0.0,0.0,1.0}
\definecolor{green}{rgb}{0.0,1.0,0.0}

\begin{document}

\title{
MOA-2022-BLG-033Lb, KMT-2023-BLG-0119Lb, and KMT-2023-BLG-1896Lb: Three low
mass-ratio microlensing planets detected through dip signals
}
\titlerunning{MOA-2022-BLG-033Lb, KMT-2023-BLG-0119Lb, and KMT-2023-BLG-1896Lb}

\author{
     Cheongho~Han\inst{\ref{inst1}} 
\and Ian~A.~Bond\inst{\ref{inst2}}
\and Youn~Kil~Jung\inst{\ref{inst3},\ref{inst22}} 
\\
(Leading authors)
\\
     Michael~D.~Albrow\inst{\ref{inst4}}   
\and Sun-Ju~Chung\inst{\ref{inst3}}      
\and Andrew~Gould\inst{\ref{inst5},\ref{inst6}}      
\and Kyu-Ha~Hwang\inst{\ref{inst3}} 
\and Chung-Uk~Lee\inst{\ref{inst3}} 
\and Yoon-Hyun~Ryu\inst{\ref{inst3}} 
\and Yossi~Shvartzvald\inst{\ref{inst7}}   
\and In-Gu~Shin\inst{\ref{inst8}} 
\and Jennifer~C.~Yee\inst{\ref{inst8}}   
\and Hongjing~Yang\inst{\ref{inst9}}     
\and Weicheng~Zang\inst{\ref{inst8},\ref{inst9}}     
\and Sang-Mok~Cha\inst{\ref{inst3},\ref{inst10}} 
\and Doeon~Kim\inst{\ref{inst1}}
\and Dong-Jin~Kim\inst{\ref{inst3}} 
\and Seung-Lee~Kim\inst{\ref{inst3}} 
\and Dong-Joo~Lee\inst{\ref{inst3}} 
\and Yongseok~Lee\inst{\ref{inst3},\ref{inst10}} 
\and Byeong-Gon~Park\inst{\ref{inst3}} 
\and Richard~W.~Pogge\inst{\ref{inst5}}
\\
(The KMTNet Collaboration)
\\
     Fumio~Abe\inst{\ref{inst11}}
\and Richard~Barry\inst{\ref{inst12}}
\and David~P.~Bennett\inst{\ref{inst12},\ref{inst13}}
\and Aparna~Bhattacharya\inst{\ref{inst12},\ref{inst13}}
\and Hirosame~Fujii\inst{\ref{inst11}}
\and Akihiko~Fukui\inst{\ref{inst14},}\inst{\ref{inst15}}
\and Ryusei~Hamada\inst{\ref{inst16}}
\and Yuki~Hirao\inst{\ref{inst16}}
\and Stela~Ishitani Silva\inst{\ref{inst12},\ref{inst17}}
\and Yoshitaka~Itow\inst{\ref{inst11}}
\and Rintaro~Kirikawa\inst{\ref{inst16}}
\and Naoki~Koshimoto\inst{\ref{inst18}}
\and Yutaka~Matsubara\inst{\ref{inst11}}
\and Shota~Miyazaki\inst{\ref{inst16}}
\and Yasushi~Muraki\inst{\ref{inst11}}
\and Greg~Olmschenk\inst{\ref{inst11}}
\and Cl{\'e}ment~Ranc\inst{\ref{inst19}}
\and Nicholas~J.~Rattenbury\inst{\ref{inst20}}
\and Yuki~Satoh\inst{\ref{inst16}}
\and Takahiro~Sumi\inst{\ref{inst16}}
\and Daisuke~Suzuki\inst{\ref{inst16}}
\and Mio~Tomoyoshi\inst{\ref{inst16}}
\and Paul~J.~Tristram\inst{\ref{inst21}}
\and Aikaterini~Vandorou\inst{\ref{inst12},\ref{inst13}}
\and Hibiki~Yama\inst{\ref{inst16}}
\and Kansuke~Yamashita\inst{\ref{inst16}}
\\
(The MOA Collaboration)
}

\institute{
      Department of Physics, Chungbuk National University, Cheongju 28644, Republic of Korea                                                          \label{inst1}     
\and  Institute of Natural and Mathematical Science, Massey University, Auckland 0745, New Zealand                                                    \label{inst2}    
\and  Korea Astronomy and Space Science Institute, Daejon 34055, Republic of Korea                                                                    \label{inst3}   
\and  University of Canterbury, Department of Physics and Astronomy, Private Bag 4800, Christchurch 8020, New Zealand                                 \label{inst4}  
\and  Department of Astronomy, Ohio State University, 140 West 18th Ave., Columbus, OH 43210, USA                                                     \label{inst5} 
\and  Max-Planck-Institute for Astronomy, K\"onigstuhl 17, 69117 Heidelberg, Germany                                                                  \label{inst6}  
\and  Department of Particle Physics and Astrophysics, Weizmann Institute of Science, Rehovot 76100, Israel                                           \label{inst7}   
\and  Center for Astrophysics $|$ Harvard \& Smithsonian 60 Garden St., Cambridge, MA 02138, USA                                                      \label{inst8}  
\and  Department of Astronomy and Tsinghua Centre for Astrophysics, Tsinghua University, Beijing 100084, China                                        \label{inst9} 
\and  School of Space Research, Kyung Hee University, Yongin, Kyeonggi 17104, Republic of Korea                                                       \label{inst10}     
\and  Institute for Space-Earth Environmental Research, Nagoya University, Nagoya 464-8601, Japan                                                     \label{inst11}     
\and  Code 667, NASA Goddard Space Flight Center, Greenbelt, MD 20771, USA                                                                            \label{inst12} 
\and  Department of Astronomy, University of Maryland, College Park, MD 20742, USA                                                                    \label{inst13}  
\and  Department of Earth and Planetary Science, Graduate School of Science, The University of Tokyo, 7-3-1 Hongo, Bunkyo-ku, Tokyo 113-0033, Japan   \label{inst14} 
\and  Instituto de Astrof{\'i}sica de Canarias, V{\'i}a L{\'a}ctea s/n, E-38205 La Laguna, Tenerife, Spain                                            \label{inst15} 
\and  Department of Earth and Space Science, Graduate School of Science, Osaka University, Toyonaka, Osaka 560-0043, Japan                            \label{inst16}  
\and  Oak Ridge Associated Universities, Oak Ridge, TN 37830, USA                                                                                     \label{inst17}   
\and  Department of Astronomy, Graduate School of Science, The University of Tokyo, 7-3-1 Hongo, Bunkyo-ku, Tokyo 113-0033, Japan                     \label{inst18}
\and  Sorbonne Universit\'e, CNRS, UMR 7095, Institut d'Astrophysique de Paris, 98 bis bd Arago, 75014 Paris, France                                  \label{inst19}
\and  Department of Physics, University of Auckland, Private Bag 92019, Auckland, New Zealand                                                         \label{inst20}    
\and  University of Canterbury Mt.~John Observatory, P.O. Box 56, Lake Tekapo 8770, New Zealand                                                       \label{inst21}  
\and  Corresponding author                                                                                                                            \label{inst22}   
}                                                                                                                                                       
\date{Received ; accepted}

\abstract
{}
{
We examined the anomalies in the light curves of the lensing events MOA-2022-BLG-033,
KMT-2023-BLG-0119, and KMT-2023-BLG-1896. These anomalies share similar traits, occurring
near the peak of moderately to highly magnified events and displaying a distinct short-term 
dip feature.
}
{
We conducted detailed modeling of the light curves to uncover the nature of the anomalies. This
modeling revealed that all signals originated from planetary companions to the primary lens. The
planet-to-host mass ratios are very low: $q\sim 7.5\times 10^{-5}$ for MOA-2022-BLG-033, $q\sim 
3.6\times 10^{-4}$ for KMT-2023-BLG-0119, and $q\sim 6.9\times 10^{-5}$ for KMT-2023-BLG-1896. 
The anomalies occurred as the source passed through the negative deviation region behind the 
central caustic along the planet-host axis. The solutions are subject to a common inner-outer 
degeneracy, resulting in variations in estimating the projected planet-host separation.  For 
KMT-2023-BLG-1896, although the planetary scenario provides the best explanation of the anomaly, 
the binary companion scenario is marginally possible.
}
{
We estimate the physical parameters of the planetary systems through Bayesian analyses based on 
the lensing observables.  While the event timescale was measured for all events, the angular 
Einstein radius was not measured for any. Additionally, the microlens parallax was measured for 
MOA-2022-BLG-033.  The analysis identifies MOA-2022-BLG-033L as a planetary system with an ice 
giant, approximately 12 times the mass of Earth, orbiting an early M dwarf star.  The companion 
of KMT-2023-BLG-1896L is also an ice giant, with a mass around 16 Earth masses, orbiting a 
mid-K-type main-sequence star.  The companion of KMT-2023-BLG-0119L, which has a mass about 
the mass of Saturn, orbits a mid-K-type dwarf star.  The lens for MOA-2022-BLG-033 is highly 
likely to be located in the disk, whereas for the other events, the probabilities of the lens 
being in the disk or the bulge are roughly comparable.	
}
{}


\maketitle

\section{Introduction} \label{sec:one}

Since 2015, the Korea Microlensing Telescope Network \citep[KMTNet;][]{Kim2016} group has been
conducting high-cadence observations of stars in the Galactic bulge region to discover exoplanets
using the microlensing method. This experiment has led to the discovery of over 3,000 lensing
events each year. About 10\% of these detected events show deviations in their light curves from
the single-lens single-source (1L1S) model due to various causes. Among these anomalous events,
around 10\% of the distortions are attributed to planets, leading to an estimated annual planet
detection rate of about 30 \citep{Gould2022b}.

\begin{table*}[t]
\caption{Coordinates, extinction, and baseline magnitude.  \label{table:one}}
\begin{tabular}{llcllll}
\hline\hline
\multicolumn{1}{c}{Event}                      &
\multicolumn{1}{c}{(RA, DEC)$_{\rm J2000}$}    &
\multicolumn{1}{c}{$(l,b)$}                    &
\multicolumn{1}{l}{$A_I$ (mag)}                &
\multicolumn{1}{c}{$I_{\rm base}$ (mag)}       &
\multicolumn{1}{c}{Other ID}                  \\
\hline
 MOA-2022-BLG-033  & (18:11:30.16, -25:03:11.09) &  $( 6^\circ\hskip-2pt .2442$, -$31^\circ\hskip-2pt .0558)$ &  1.88   &  18.43 &  KMT-2022-BLG-0118   \\
 KMT-2023-BLG-0119 & (17:47:15.61, -34:34:19.88) &  (-$4^\circ\hskip-2pt .6398$, -$31^\circ\hskip-2pt .2220)$ &  1.37   &  20.61 &  MOA 2023-BLG-104    \\
 KMT-2023-BLG-1896 & (18:04:07.62, -26:57:32.11) &  $( 3^\circ\hskip-2pt .7714$, -$21^\circ\hskip-2pt .5289)$ &  1.33   &  19.91 &                      \\
\hline
\end{tabular}
\end{table*}

In gravitational lensing events, planetary signals manifest in diverse forms within the light curve.
These signals arise as a planet produces a caustic on the source plane. The size and shape of the
caustic created by the planet vary depending on the separation between the planet and its host
star, as well as the mass ratio between them. Additionally, the planetary signal displays a wide
range of forms depending on the source path relative to the caustic.

As the number of discovered planets from the KMTNet survey grows, those with similar signal
characteristics are now being grouped and reported collectively. \citet{Han2024b} identified 
a distinctive pattern of planetary signals through their analysis of four lensing events:
KMT-2020-BLG-0757, KMT-2022-BLG-0732, KMT-2022-BLG-1787, and KMT-2022-BLG-1852. The
anomalies in the light curves of these events exhibited a shared pattern, displaying an extended
trough followed by a bump. They determined that these anomalies were produced by planets
located within the Einstein rings of their host stars. The bump in the light curve occurred when
the source star crossed a planetary caustic, while the troughs resulted from the source moving
through a region of minor image perturbations situated between a pair planetary caustics.

Planetary signals generated by the source approaching peripheral caustics created by planets
outside the Einstein ring exhibit distinct characteristics. Due to the location of the caustic, 
these signals typically appear in the wings of the lensing light curve. Characteristic features 
of such planetary signals were demonstrated by \citet{Jung2021} for the events OGLE-2018-BLG-0567
and OGLE-2018-BLG-0962, and by \citet{Han2024c} for KMT-2021-BLG-2609 and
KMT-2022-BLG-0303.

\citet{Han2024a} identified another distinctive pattern in planetary signals through their 
analysis of events MOA-2022-BLG-563, KMT-2023-BLG-0469, and KMT-2023-BLG-0735. These anomalies
share a characteristic feature, with each anomaly occurring near the peak of a high-magnification
event. The central part of the anomaly displays a dip, flanked by subtle bumps on both sides.
They found that interpreting these anomalies involves a common inner-outer degeneracy, leading 
to
ambiguity in estimating the projected separation between the planet and its host.

\citet{Han2021b} illustrated planetary signals induced by giant planets near the Einstein 
ring in events KMT-2017-BLG-2509, OGLE-2017-BLG-1099, and OGLE-2019-BLG-0299. They noted 
that, due to the large size of the caustic, the duration of these planetary signals constitutes 
a significant portion of the total event, making it challenging to readily identify their 
planetary nature.

Planetary signals can be produced when the source approaches the caustic without crossing it.
Examples of such signals, which lack caustic-crossing features, were presented by \citet{Han2023b} 
for lensing events KMT-2022-BLG-0475 and KMT-2022-BLG-1480, and by \citet{Han2021a} for events 
KMT-2018-BLG-1976, KMT-2018-BLG-1996, and OGLE-2019-BLG-0954. \citet{Han2022} illustrated examples 
of well-covered planetary signals resulting from high-cadence observations and data integration 
across multiple surveys for events OGLE-2017-BLG-1691, KMT-2021-BLG-0320, KMT-2021-BLG-1303, 
and KMT-2021-BLG-1554. In contrast, examples of planets identified from partially covered signals 
were provided by \citet{Han2023a} for events KMT-2018-BLG-1976, KMT-2018-BLG-1996, and 
OGLE-2019-BLG-0954.

In this paper, we present the discovery of three microlensing planets with similar signal
characteristics, identified through the analysis of the lensing events MOA-2022-BLG-033,
KMT-2023-BLG-0119, and KMT-2023-BLG-1896. The planetary signals share a common trait in
which they appear near the peak of moderately to highly magnified events, exhibiting a 
distinct short-term dip feature.

\section{Observation and data} \label{sec:two}

In Table~\ref{table:one}, we list the equatorial and Galactic coordinates, $I$-band 
extinction ($A_I$), and baseline magnitude ($I_{\rm base}$) for the lensing events 
MOA-2022-BLG-033, KMT-2023-BLG-0119, and KMT-2023-BLG-1896.  The extinction is estimated 
as $A_I = 7A_K$, where the $K$-band extinction is adopted from \citet{Gonzalez2012}.  While 
the anomalies in their lensing light curves were initially detected by examining KMTNet data, 
we later found that both MOA-2022-BLG-033 and KMT-2023-BLG-0119 were also observed by the 
Microlensing Observations in Astrophysics \citep[MOA][]{Bond2001, Sumi2003} group.  In the 
table, we include the ID references assigned by each group for these events.  For events 
identified by both groups, we use the ID reference of the group that first discovered them.

The KMTNet survey is conducted using three identical telescopes, each with a 1.8-meter aperture
and equipped with cameras that cover 4 square degrees of the sky. These telescopes are located 
at Siding Spring Observatory in Australia (KMTA), Cerro Tololo Interamerican Observatory in Chile
(KMTC), and South African Astronomical Observatory in South Africa (KMTS). The MOA group uses 
a 1.8-meter telescope equipped with a camera that covers 2.2 square degrees, located at Mt. 
John University Observatory in New Zealand.

The KMTNet survey mainly observed in the $I$-band, whereas the MOA survey utilized a specially 
designed MOA-$R$ band, spanning wavelengths from 609 to 1109 nm. Photometry for each microlensing 
event was carried out using software developed by each survey group. The KMTNet group utilized 
a photometry code developed by \citet{Albrow2009}, while the MOA survey employed a code created 
by \citet{Bond2001}.  Both codes were developed based on the difference imaging method 
\citep{Tomaney1996, Alard1998}.  For the KMTNet data, we conducted additional photometry using 
the code developed by \citet{Yang2024} to ensure optimal data quality.  We adjusted the error 
bars to make them consistent with the data scatter and to set the value of $\chi^2$ per degree 
of freedom to unity for each data set, following the procedure outlined in \citet{Yee2012}.

\begin{table*}[t]
\caption{Lensing parameters of MOA-2022-BLG-033.  \label{table:two}}
\begin{tabular}{lllllllll}
\hline\hline
\multicolumn{1}{c}{Parameter}  &
\multicolumn{2}{c}{Inner}      &
\multicolumn{2}{c}{Outer}      \\
\multicolumn{1}{c}{}           &
\multicolumn{1}{c}{$u_0>0$}    &
\multicolumn{1}{c}{$u_0<0$ }   &
\multicolumn{1}{c}{$u_0>0$ }   &
\multicolumn{1}{c}{$u_0<0$ }   \\
\hline
  $\chi^2$                &   $2649.5            $    &  $2652.5            $  &   $2650.7            $     &   $2653.5            $    \\
  $t_0$ (HJD$^\prime$)    &   $9658.383 \pm 0.063$    &  $9658.368 \pm 0.066$  &   $9658.387 \pm 0.064$     &   $9658.347 \pm 0.064$    \\
  $u_0$                   &   $0.1225 \pm 0.0045 $    &  $-0.1145 \pm 0.0036$  &   $0.1237 \pm 0.0046 $     &   $-0.1179 \pm 0.0035$    \\
  $\te$ (days)            &   $113.85 \pm 3.08   $    &  $122.33 \pm 3.17   $  &   $112.39 \pm 3.05   $     &   $119.68 \pm 3.10   $    \\
  $s$                     &   $0.9041 \pm 0.0058 $    &  $0.9099 \pm 0.0055 $  &   $0.9770 \pm 0.0060 $     &   $0.9739 \pm 0.0059 $    \\
  $q$ (10$^{-5}$)         &   $7.26 \pm  1.29    $    &  $6.59 \pm 1.14     $  &   $8.20 \pm 1.26     $     &   $6.85 \pm 1.20     $    \\
  $\alpha$ (rad)          &   $1.3042 \pm 0.0048 $    &  $-1.3039 \pm 0.0046$  &   $1.3016 \pm 0.0049 $     &   $-1.3047 \pm 0.0048$    \\
  $\rho$ (10$^{-3}$)      &   $< 6               $    &  $ < 6              $  &   $< 6               $     &   $ < 6              $    \\
  $\pien$                 &   $-0.293 \pm 0.049  $    &  $0.367 \pm 0.052   $  &   $-0.297 \pm  0.050 $     &   $0.375 \pm 0.049   $    \\		  
  $\piee$                 &   $0.134 \pm 0.018   $    &  $0.137 \pm 0.016   $  &   $0.143 \pm 0.017   $     &   $0.145 \pm 0.016   $    \\		  
\hline
\end{tabular}
\end{table*}

\section{Modeling procedure} \label{sec:three}

A planetary signal in the light curve of a lensing event occurs when the source approaches 
the caustic generated by a planet companion to the lens \citep{Mao1991, Gould1992}.  Caustics 
indicate the positions on the source plane where lensing causes the magnification of a point 
source to become infinite. The planet creates two sets of caustics. The first set forms around 
the host star (central caustic), while the second set forms at approximately the position of 
${\bf s}-1/{\bf s}$ from the host star (planetary caustic). Here, ${\bf s}$ represents the 
position vector of the planet with respect to its host, with its length scaled to the angular 
Einstein radius ($\thetae$) of the lens \citep{Chung2005, Han2006}. The central caustic forms 
a closed, wedge-shaped curve made up of concave segments. Because the central caustic is 
located near the primary lens, planetary signals caused by it appear near the peak of a high 
magnification event \citep{Griest1998}. While the strongest planetary signal occurs when the 
source crosses the caustic, signals can also be generated without crossing it.\footnote{
\citet{Zhu2014} predicted that half of all planets detected in a KMTNet-like survey would lack 
caustic crossings, and \citet{Jung2023} confirmed this prediction for a complete sample of 
2018+2019 KMTNet planets in their Table 17. \citet{Gould2022a} showed that roughly one third of 
planetary events that lack caustic crossings nevertheless show sufficiently strong finite-source 
effects to yield estimates of the the normalized source radius $\rho=\theta_*/\theta_{\rm E}$, 
and therefore of the Einstein radius, $\theta_{\rm E}$.} The source passing through the region 
in front of the protruding cusp of the central caustic shows positive deviations, whereas passing 
through the back region of the caustic results in negative deviations.

The light curves of the three analyzed lensing events show the following common characteristics.
First, a very short-duration anomaly appears in the lensing light curve. Second, this anomaly
occurs near the peak of the light curve with moderate to high magnification. Third, the anomaly
exhibits a smooth light variation without abrupt changes. Fourth, the signal shows a negative
deviation relative to the underlying 1L1S light curve of the event. These characteristics 
suggest that the anomaly is very likely of planetary origin and is produced by the source 
passing through the back region of the central caustic induced by the planet.  A short-term 
anomaly can be caused by a faint companion to the source \citep{Gaudi1998}. However, because a 
binary source induces only positive deviations, we rule out the binary source as the origin of 
the anomaly.

Given the likely planetary origin of the observed anomalies, we conducted a binary-lens 
single-source (2L1S) modeling of the light curves. To represent the light curve of a 2L1S event, 
seven basic parameters are required. The first three parameters $(t_0, u_0, \te)$ describe the 
approach of the lens and source, where each parameter represents the time of closest approach, 
the projected separation at that time (scaled to $\thetae$), and the event timescale, respectively. 
The next two parameters $(s, q)$ describe the binary lens, with each parameter representing the 
projected separation between the two lens components ($M_1$ and $M_2$) and their mass ratio.
Here the separation $s$ is scaled to $\thetae$. An additional parameter, $\alpha$, denotes the 
angle between the source's motion vector and the binary lens axis. The final parameter, $\rho$, 
represents the ratio of the angular source radius to the angular Einstein radius ($\rho=\theta_*
/\thetae$, normalized source radius) and is necessary to describe the deformation of the light 
curve due to the finite-source effect when the source crosses (or comes very close to) the caustic.

In addition to the basic parameters, additional parameters are needed for certain special cases 
of events.  One such case is an event with a very long timescale. In this case, the relative 
lens-source motion experiences acceleration due to the parallactic motion induced by the Earth's 
orbit around the Sun \citep{Gould1992, Gould2000, Gould2004}. Consequently, additional parameters 
$(\pien, \piee)$ should be included in the modeling. These parameters represent the north and east 
components of the microlens-parallax vector, defined as
\begin{equation}
\pie =  
\left( {\pi_{\rm rel} \over \thetae}\right)
\left( {\muvec \over \mu}\right),
\label{eq1}
\end{equation}
where $\pi_{\rm rel} ={\rm AU}(\dl^{-1}-\ds^{-1})$ is the relative lens-source parallax, $\dl$ 
and $\ds$ denote the distances to the lens and source, respectively, and $\muvec$ denotes the 
relative lens-source proper motion vector.

In the modeling, we searched for the lensing parameters that best describe the anomaly. To find
the binary parameters $(s, q)$, we used a grid approach because the lensing magnification varies
discontinuously with changes in these parameters. For the other parameters, for which the 
magnification varies smoothly with parameter changes, we used a downhill approach with multiple 
starting values of $\alpha$. For the downhill approach, we employed the Markov chain Monte Carlo 
(MCMC) method.  For each individual local solution identified in the $\chi^2$ map on the $s$--$q$ 
parameter plane, we refined the solution by allowing all parameters to vary.

It is known that there is degeneracy in the interpretation of planetary signals in lensing 
light curves. The most well-known two types are the "inner-outer" degeneracy and the 
"close-wide" degeneracy. The former degeneracy arises because the light curves resulting 
from a source passing through the inner and outer regions of the planetary caustic produce 
similar anomalies \citep{Gaudi1997}. The latter degeneracy occurs because the central caustics 
created by planets with $s<1$ (close) and $s>1$ (wide) generate similar anomalies 
\citep{Griest1998}. Building on the work of \citet{Herrera2020}, \citet{Yee2021}, 
\citet{Hwang2022}, \citet{Zhang2022}, and \citet{Gould2022b}, it is now established that the 
two types of degeneracy can be unified, with an analytic expression developed to describe 
the relationships between lensing parameters for solutions affected by both types of degeneracy.  
Hereafter, we use the unified term "inner-outer" to both types of degeneracy arising in the 
interpretation of planetary signals.  The relationship between the planet separations of the 
inner ($s_{\rm in}$) and outer ($s_{\rm out}$) solutions under this degeneracy is expressed as
\begin{equation}
\left( s_{\rm in} \times s_{\rm out} \right)^{1/2}
= s^\dagger;\qquad
s^\dagger = 
{\sqrt{u_{\rm anom}^2+4} \pm u_{\rm anom}\over 2},
\label{eq2}
\end{equation}
where $u_{\rm anom}= (\tau_{\rm anom}^2+u_0^2)^{1/2}$, $\tau_{\rm anom}=(t_{\rm anom}-t_0)/\te$, 
and $t_{\rm anom}$ denotes the time of the planet-induced anomaly. The signs "$+$" and "$-$" 
in the second term apply to anomalies with bump and dip features, respectively. Because all 
the anomalies in the analyzed events exhibit dip features, the sign is "$-$". In our analysis 
of the events, we examine whether the observed anomalies are subject to this degeneracy.

\begin{figure}[t]
\includegraphics[width=\columnwidth]{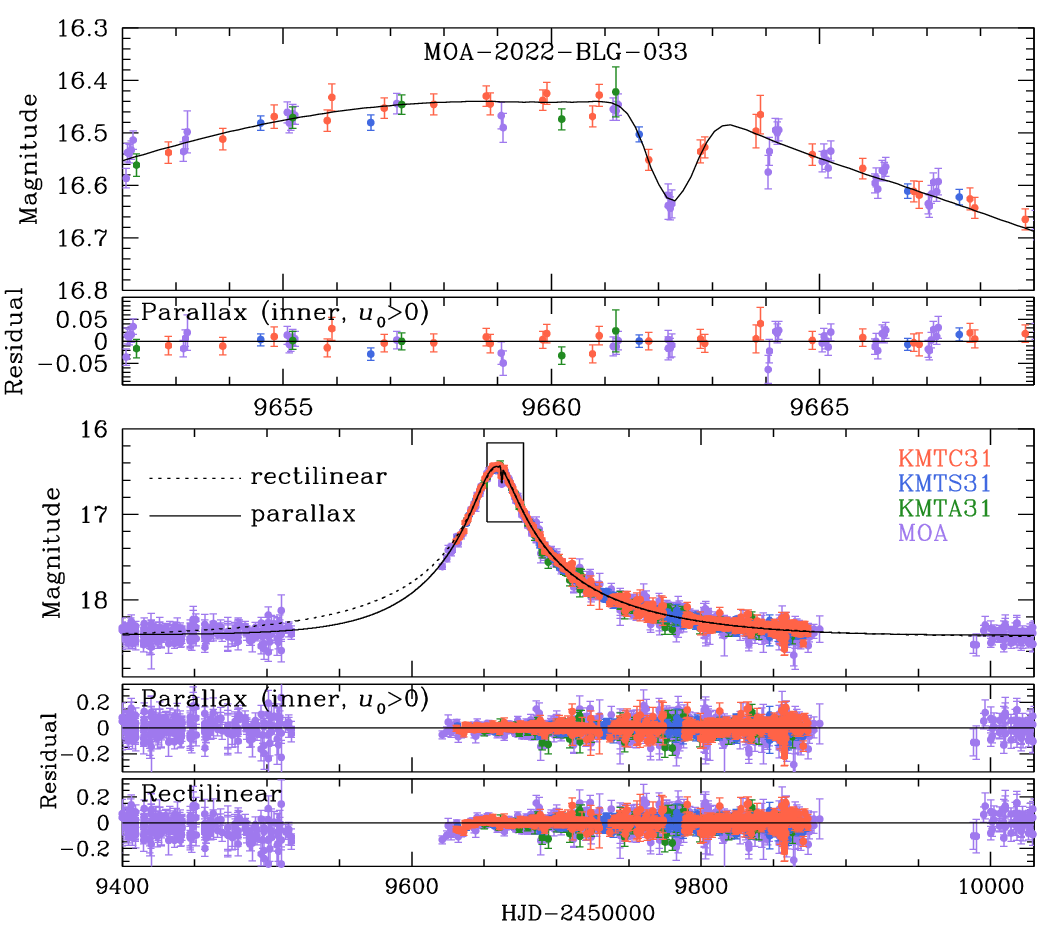}
\caption{
Light curve the lensing event MOA-2022-BLG-033 and the model curve.  The lower panel offers a 
comprehensive view of the event, while the upper panel provides a close-up view of the anomaly 
(the region enclosed by a box in the lower panels).  The colors of the data points correspond 
to the telescopes listed in the legend.  The bottom two panels show the residuals from the 
models with and without considering microlens-parallax effects.
}
\label{fig:one}
\end{figure}

\section{Analysis} \label{sec:four}

In this section, we present the analyses conducted for each individual event and their 
corresponding results. For each event, we begin with a brief description of its discovery and 
the anomaly observed in the light curve. We then present the lensing solutions derived from 
the modeling, along with a discussion of any degeneracies that arise in the interpretation.

\subsection{MOA-2022-BLG-033} \label{sec:four-one}

The lensing event MOA-2022-BLG-033 was initially detected by the MOA group on February 20,
2022, which corresponds to the abridged Heliocentric Julian date ${\rm HJD}^\prime \equiv 
{\rm HJD}-2450000=9630$. About a month later, the KMTNet group confirmed the event on March 
23 (${\rm HJD}^\prime=9661$). The event had a long duration, with the lensing magnification 
starting before the 2022 season began and continuing until the season ended.  In our analysis, 
we include the MOA data from the 2021 and 2023 seasons to ensure accurate measurement of the 
baseline magnitude.  The source of the event lies in the KMTNet BLG31 field toward which 
observations were conducted with a 2.5-hour cadence.

\begin{figure}[t]
\includegraphics[width=\columnwidth]{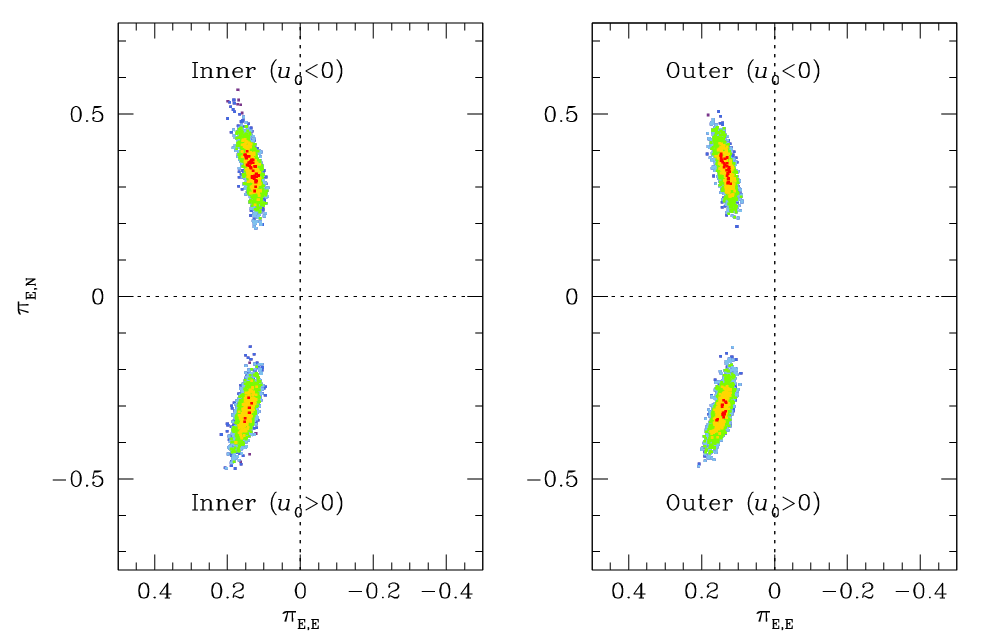}
\caption{
Scatter plots of points in the MCMC chain on the $(\piee,\pien)$ parameter plane.  Colors are chosen 
to present points with $\leq 1\sigma$ (red), $\leq 2\sigma$ (yellow), $\leq 3\sigma$ (green), 
$\leq 4\sigma$ (cyan), and $\leq 5\sigma$ (blue).
}
\label{fig:two}
\end{figure}

\begin{figure}[t]
\includegraphics[width=\columnwidth]{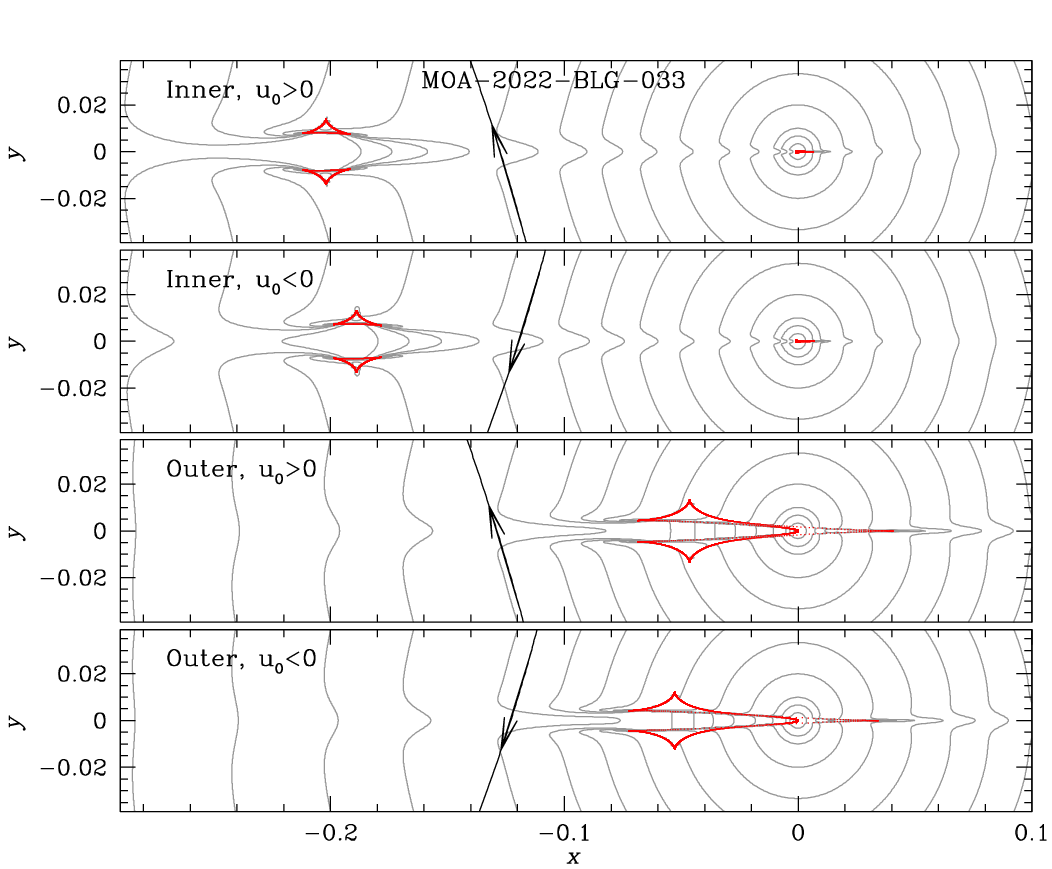}
\caption{
Lens-system configurations for the four degenerate solutions of MOA-2022-BLG-033. In each 
panel, the red figure represents the caustic, and the arrowed curve indicates the source
trajectory. The grey curves surrounding the caustic represent equi-magnification contours.
The coordinates are centered on the position of the primary lens, and the lengths are scaled 
to the Einstein radius.
}
\label{fig:three}
\end{figure}

Figure~\ref{fig:one} shows the light curve of the event. It appears to be a typical 1L1S event 
with a moderately high magnification of $A_{\rm max}\sim 7.9$ and a very long timescale of 
$\te \sim 110$~days.  A close examination of the peak region revealed a very short-term anomaly.  
The zoomed-in view of the region around the anomaly is presented in the upper panel. The anomaly, 
covered by the MOA, KMTC, KMTS data sets, exhibited a negative deviation from the 1L1S model 
with a duration of about $\Delta t_{\rm anom}\sim 2$~days.  The pattern and duration of the 
anomaly suggest it was caused by a planetary companion to the lens.

\begin{figure}[t]
\includegraphics[width=\columnwidth]{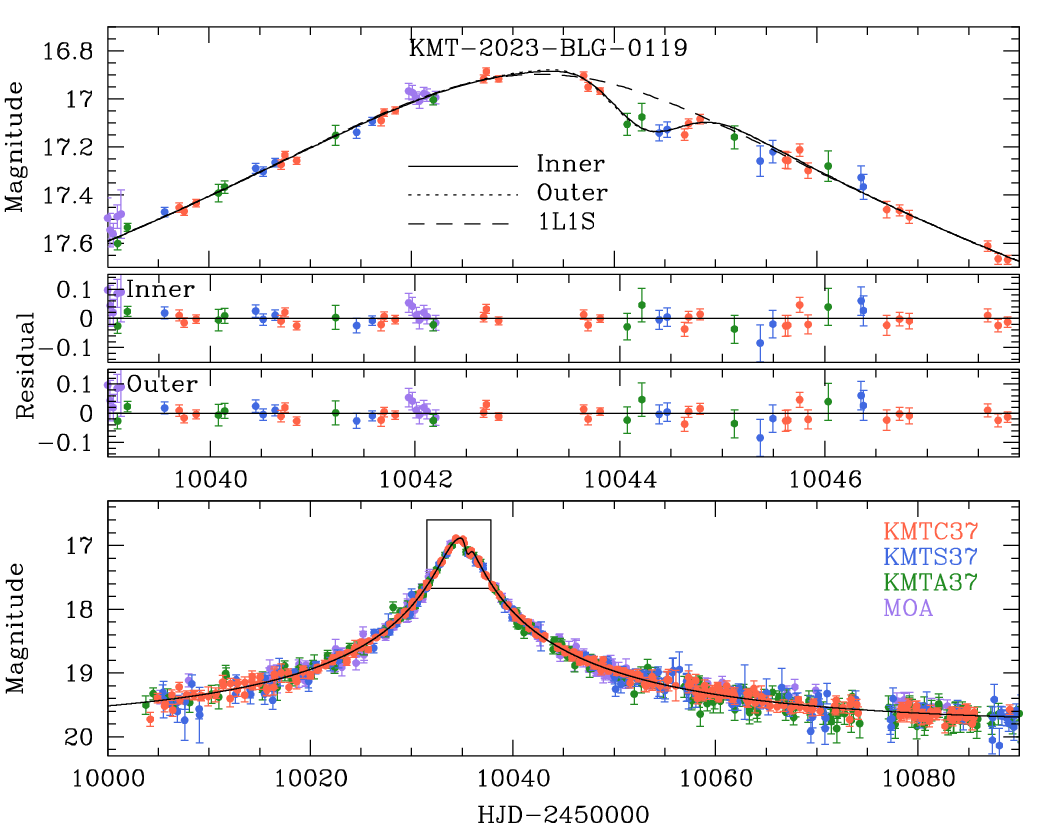}
\caption{
Light curve of KMT-2023-BLG-0119.  The second and third panels display the residuals for the 
inner and outer solutions, respectively.
}
\label{fig:four}
\end{figure}

Detailed modeling confirmed the planetary nature of the anomaly. In Table~\ref{table:two}, we 
list the lensing parameters of the solutions.  We present the modeling solutions that account 
for microlens-parallax effects, with which the fit improves by $\Delta\chi^2 \sim 290$ compared 
to the model assuming rectilinear lens-source motion.  We identified four sets of solutions due 
to two types of degeneracies. First, the event displayed an inner-outer degeneracy between 
solutions with planetary separations of approximately $s_{\rm in} \sim 0.90$ and $s_{\rm out} 
\sim 0.97$.  The second type of degeneracy involved parallax solutions with $u_0 > 0$ and $u_0 
< 0$, resulting from the mirror symmetry of the source trajectory relative to the planet-host 
axis. \citep{Smith2003, Skowron2011}.  Figure~\ref{fig:two} shows the scatter plots of points 
in the MCMC chain on the $(\piee, \pien)$ plane for the four sets of solutions.  
The degeneracy among the solutions are very severe with $\Delta\chi^2 < 4$.
The model curve and residuals for the inner solution with $u_0>0$ are shown in the figure.  
We note that the model curves of the other solutions are very similar to the one presented, and 
thus they are not displayed. To illustrate the parallax effect, we also present the model curve 
and residuals for the solution obtained under the assumption of rectilinear lens-source motion.  
As anticipated from the very small duration ratio of the anomaly to the event, the planet-to-host 
mass ratio, $q \sim 0.71 \times 10^{-4}$, is very low. The mass ratios estimated for the individual 
solutions are consistent across all cases.  Although the exact value of the normalized source 
radius could not be determined because of the anomaly not crossing the caustic, an upper limit 
of $\rho_{\rm max}\sim 6\times 10^{-3}$ can be established.

Figure~\ref{fig:three} illustrates the lens-system configurations for the four degenerate 
solutions of MOA-2022-BLG-033.  For the inner solution, the lens produces two distinct sets 
of central and planetary caustics, with the source passing through the region between them.  
In contrast, the outer solution features a single set of resonant caustics in which the central 
and planetary caustics merge, and the source traversed the outer region of the caustic.  In 
both cases, the anomaly occurred as the source traversed the negative deviation region along 
the planet-host axis.  From the lensing parameters $(t_0, u_0, \te, t_{\rm anom}) \sim (9658.4, 
0.123, 113, 9662.2)$, we find $s^\dagger =0.938$.  This matches very well the geometric mean 
$(s_{\rm in}\times s_{\rm out})^{1/2} =0.940$, indicating that the pair of inner-outer solutions 
well follow the formalism in Eq.~(\ref{eq2}).  The parameters for the solutions with $u_0 > 0$ 
and $u_0 < 0$ are approximately related as $(u_0, \alpha, \pien)_{u_0 > 0} \leftrightarrow -
(u_0, \alpha, \pien)_{u_0 < 0}$.

\subsection{KMT-2023-BLG-0119} \label{sec:four-two}

The KMTNet group initially detected the lensing event KMT-2023-BLG-0119 on March 20, 2023
(${\rm HJD}^\prime =10023$). The MOA group confirmed the event on April 4 (${\rm HJD}^\prime 
= 10038$) and designates it as MOA 2023-BLG-104. Similar to the previous event, the magnification 
of the source flux started before the 2023 season. The source of the event is located in the 
KMTNet BLG37 field, with observations made at a 2.5-hour cadence.

Figure~\ref{fig:four} shows the light curve of the event, constructed from the combined 
KMTNet and MOA data. At first glance, it appears to be a typical 1L1S event with a moderately 
high magnification of $A_{\rm max}\sim 23.4$. However, a careful examination of the peak region 
revealed a short-term anomaly lasting about $\Delta t_{\rm anom}\sim 1.5$~days. The upper panel 
provides an enlarged view of the region around the anomaly.  Despite its short duration, the 
anomaly was captured by the combined data from all three KMTNet sets.  This anomaly displays 
a negative deviation from the underlying 1L1S curve.  The characteristics of this anomaly are 
very similar to those of MOA-2022-BLG-033, suggesting a planetary origin.

\begin{figure}[t]
\includegraphics[width=\columnwidth]{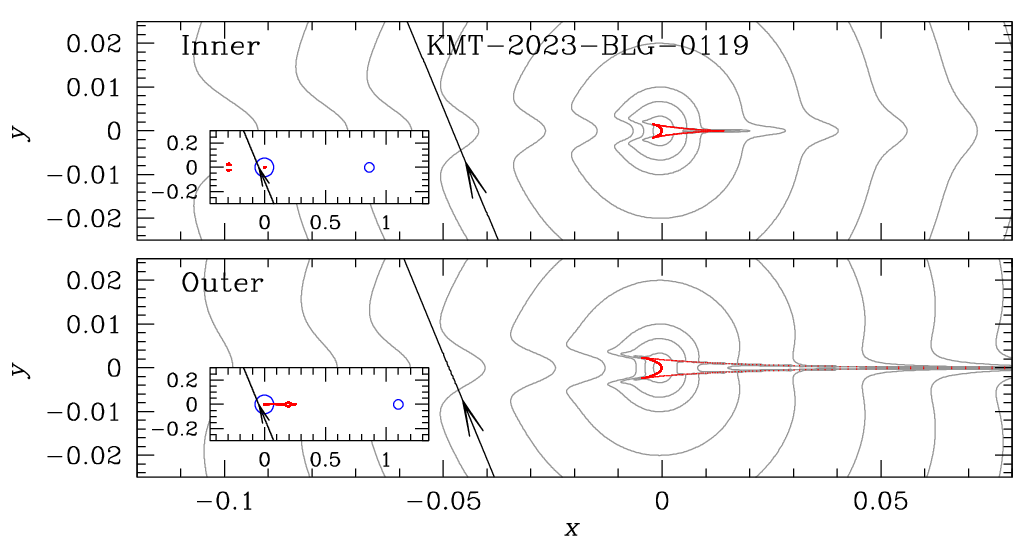}
\caption{
Configurations of the lens system for the inner and outer solutions of KMT-2023-BLG-0119.
The inset in each panel offers a zoomed-out view. In the inset, blue empty circles mark the 
positions of the lens components: a small circle for the planet and a large circle for the host.
}
\label{fig:five}
\end{figure}

\begin{table}[t]
\caption{Lensing parameters of KMT-2023-BLG-0119.\label{table:three}}
\begin{tabular*}{\columnwidth}{@{\extracolsep{\fill}}lllll}
\hline\hline
\multicolumn{1}{c}{Parameter}    &
\multicolumn{1}{c}{Inner}        &
\multicolumn{1}{c}{Outer}        \\
\hline
  $\chi^2$               &  $1172.28            $  &  $1171.90            $  \\ 
  $t_0$ (HJD$^\prime$)   &  $10043.242 \pm 0.015$  &  $10043.240 \pm 0.015$  \\   
  $u_0$ (10$^{-2}$)      &  $4.41 \pm 0.19      $  &  $4.50 \pm 0.19      $  \\      
  $\te$ (days)           &  $58.95 \pm 2.16     $  &  $57.90 \pm 2.19     $  \\       
  $s$                    &  $0.862 \pm 0.040    $  &  $1.101 \pm 0.050    $  \\     
  $q$  (10$^{-4}$)       &  $3.547 \pm 1.10     $  &  $3.729 \pm 0.98     $  \\   
  $\alpha$ (rad)         &  $1.177 \pm 0.017    $  &  $1.179 \pm 0.014    $  \\      
  $\rho$ (10$^{-3}$)     &  $< 10               $  &  $< 10               $  \\
\hline                                                   
\end{tabular*}
\end{table}

We confirm the planetary origin of the anomaly from the detailed modeling of the light curve. 
We identify a pair of solutions resulting from the inner--outer degeneracy. In Table~\ref{table:three}, 
we list the lensing parameters of the two solutions together with the $\chi^2$ values of the fits. 
The degeneracy between the solutions is found to be very severe, with the outer solution being 
favored by only $\Delta\chi^2 =0.38$. The planetary parameters are $(s, q)_{\rm in}\sim (0.86, 
3.5\times 10^{-4})$ for the inner solution and $(s, q)_{\rm out}\sim (1.10, 3.7\times 10^{-4})$ 
for the outer solution. The model curves of both solutions are presented in Figure~\ref{fig:four}. 
The value $s^\dagger =0.976,$ derived from the estimated lensing parameters $(t_0, u_0, \te, 
t_{\rm anom}) \sim (10043.2, 4.4\times 10^{-2}, 58, 10044.2)$, is very close to the geometric mean 
of $(s_{\rm in} \times s_{\rm out})^{1/2} =0.974$.  This indicates that the degeneracy between the 
solutions arises from the inner--outer degeneracy.  The weak finite-source effects set a loose upper 
limit on the normalized source radius at $\rho_{\rm max}\sim 10^{-2}$.  Determining the microlens 
parallax parameters proved challenging due to the relatively large photometric uncertainties in 
the data.

\begin{table*}[t]
\caption{Lensing parameters of KMT-2023-BLG-1896. \label{table:four}}
\begin{tabular}{lllllllll}
\hline\hline
\multicolumn{1}{c}{Parameter}  &
\multicolumn{2}{c}{Planet}     &
\multicolumn{2}{c}{Binary}    \\
\multicolumn{1}{c}{}           &
\multicolumn{1}{c}{Inner}      &
\multicolumn{1}{c}{Outer}      &
\multicolumn{1}{c}{Inner}      &
\multicolumn{1}{c}{Outer}     \\
\hline
  $\chi^2$                &   $4786.04                        $   &  $4785.78                      $  &  $4795.96            $    &  $4795.68            $  \\
  $t_0$ (HJD$^\prime$)    &   $10159.1012 \pm 0.0030          $   &  $10159.1020 \pm 0.0029        $  &  $159.1039 \pm 0.0014$    &  $159.0994 \pm 0.0032$  \\
  $u_0$ (10$^{-3}$)       &   $1.94 \pm 0.62                  $   &  $1.56 \pm 0.59                $  &  $2.64 \pm 0.54      $    &  $2.21 \pm 0.57      $  \\
  $\te$ (days)            &   $54.03 \pm 14.83                $   &  $67.10 \pm 15.92              $  &  $43.48 \pm 8.48     $    &  $50.91 \pm 9.52     $  \\
  $s$                     &   $0.786 \pm 0.048                $   &  $1.268 \pm 0.080              $  &  $0.097 \pm 0.028    $    &  $10.42 \pm 3.77     $  \\
  $q$                     &   $(8.31 \pm 3.97) \times 10^{-5} $   &  $(6.86 \pm 4.29)\times 10^{-5}$  &  $0.100 \pm 0.092    $    &  $0.075 \pm 0.158    $  \\
  $\alpha$ (rad)          &   $2.235 \pm 0.023                $   &  $2.237 \pm 0.023              $  &  $-0.237 \pm 0.039   $    &  $-0.233 \pm 0.044   $  \\
  $\rho$ (10$^{-3}$)      &   $<1                             $   &  $<1                           $  &  $<3                 $    &  $<3                 $  \\
\hline
\end{tabular}
\end{table*}

The lens system configurations corresponding to the inner and outer solutions are presented 
in Figure~\ref{fig:five}.  For the inner solution, the source passed through the region between 
the central and planetary caustics, while for the outer solution, it traversed the outer region 
of the caustic.

\subsection{KMT-2023-BLG-1896} \label{sec:four-three}

The event KMT-2023-BLG-1896 was detected on August 4, 2023 (${\rm HJD}^\prime =10160$) and 
was exclusively observed by the KMTNet group.  The source is located in the KMTNet BLG03 
field, which was monitored with a 0.5-hour cadence.  Most of this field overlaps with the 
BLG43 field, but the source's position falls within a narrow strip that does not overlap, 
resulting in no data from the BLG43 field.  The maximum magnification at the peak was 
extremely high, reaching approximately $A_{\rm max}\sim 690$.

\begin{figure}[t]
\includegraphics[width=\columnwidth]{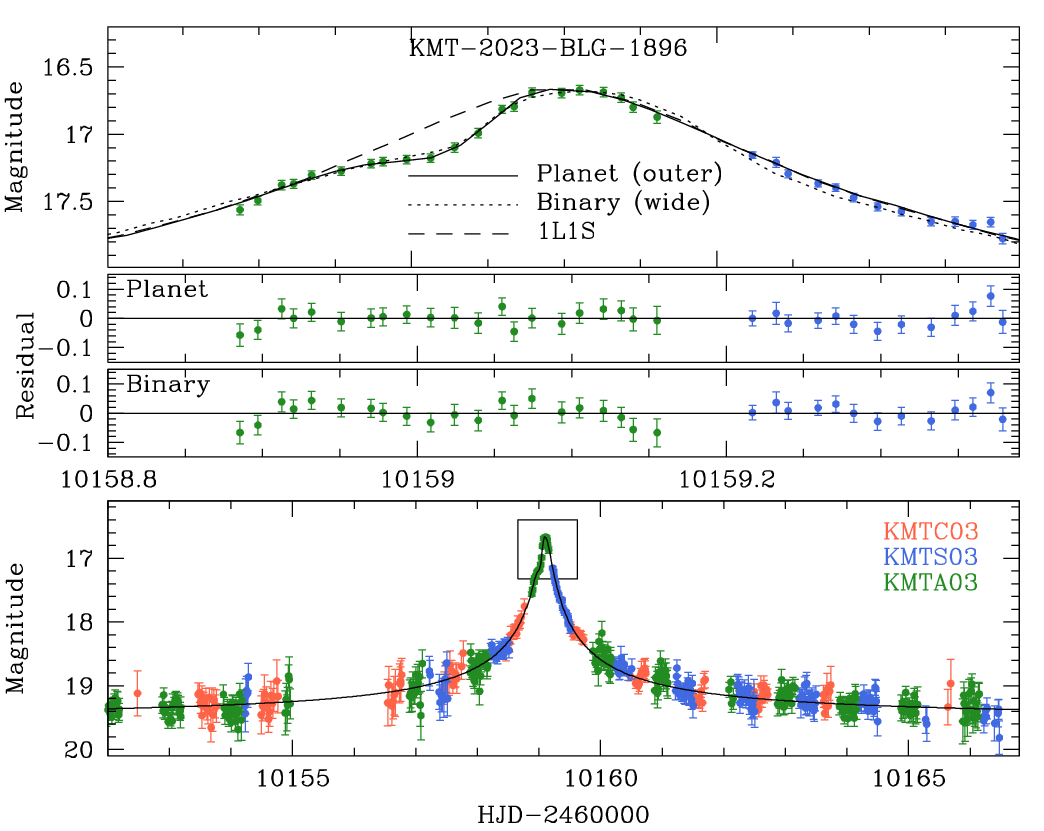}
\caption{
Light curve of KMT-2023-BLG-1896.	
}
\label{fig:six}
\end{figure}

Figure~\ref{fig:six} presents the light curve for KMT-2023-BLG-1896. As with the earlier
events, it features a brief central anomaly, lasting approximately $\Delta t_{\rm anom}
\sim 4.5$~hours, and showing a negative deviation from the baseline 1L1S model. These 
features of the anomaly suggest a planetary origin. Despite its brief duration, the anomaly 
was well-captured by the KMTA data set due to the relatively high cadence of the observations 
in the field.

Interpreting the anomaly in KMT-2023-BLG-1896 was subject to two types of degeneracy. The 
first type is the inner--outer degeneracy, which has been observed in previous events. 
The second type is the planet--binary degeneracy, which was not seen in earlier events, 
although it is relatively less severe.  In Table~\ref{table:four}, we list four sets of 
solutions, in which the first pair corresponds to the inner and outer planetary solutions, 
while the other pair corresponds to the close and wide binary solutions.  In the planetary 
interpretation, the binary parameters are $(s, q)_{\rm in} \sim (0.79, 8.3 \times 10^{-5})$ 
for the inner solution and $(s, q)_{\rm out} \sim (1.27, 6.9 \times 10^{-5})$ for the outer 
solution.  This suggests that the lens is a planetary system with the planet positioned near 
the Einstein ring of the host star.  In the binary interpretation, the parameters are 
$(s, q)_{\rm close} \sim (0.10, 0.1)$ for the inner solution and $(s, q)_{\rm wide}\sim (10.4, 
0.08)$ for the wide solution, indicating that the lens system consists of two stars with a 
projected separation significantly smaller or larger than the Einstein radius. The degeneracy 
between the solutions of each pair of solution is severe. The model curves of the outer-planetary 
and wide-binary solutions are presented in Figure~\ref{fig:six}.  It is found that the planetary 
solution provides a better fit than the binary solution, with $\Delta\chi^2 \sim 10$, as 
illustrated by the residuals in the second and third panels.  Therefore, we proceed with further 
analysis based on the planetary solution, though we cannot definitively rule out the binary 
interpretation.  The upper limit for the normalized source radius is set at $\rho_{\rm max}\sim 
10^{-3}$.  The microlens-parallax parameters could not be constrained because of the large 
uncertainty in the data resulting from the faintness of the source.

\begin{figure}[t]
\includegraphics[width=\columnwidth]{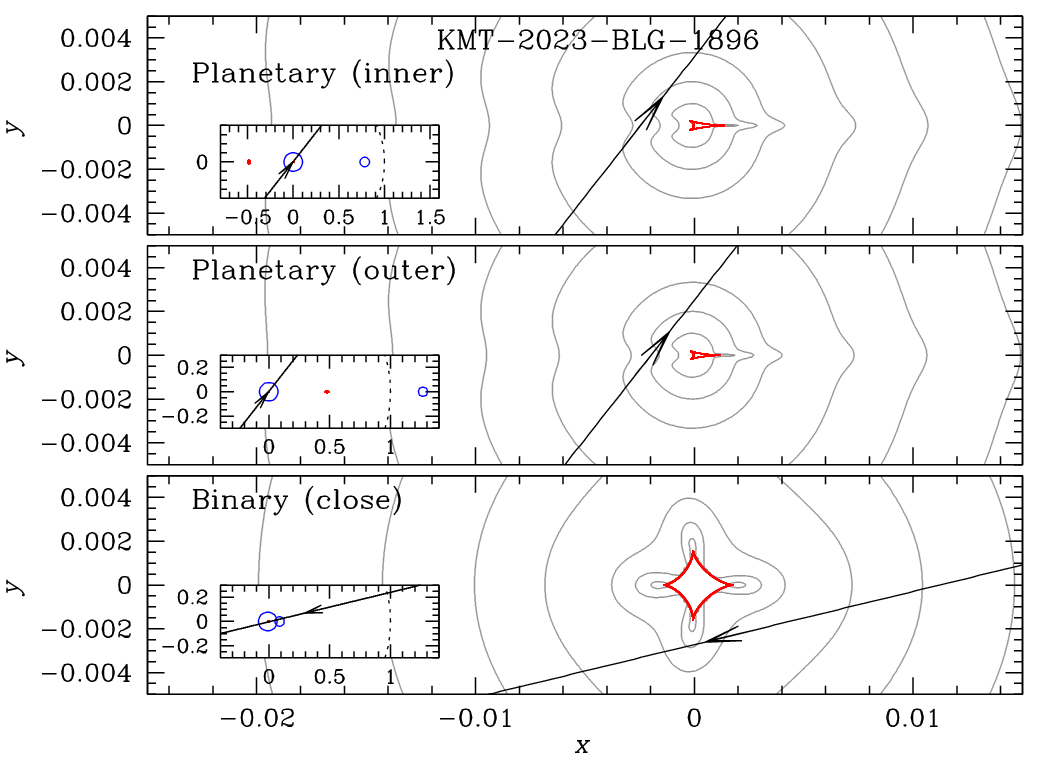}
\caption{
Lens system configurations of KMT-2023-BLG-1896 for the inner and outer
planetary solutions (upper two panels) and the wide binary solution.
}
\label{fig:seven}
\end{figure}

\begin{table*}[t]
\caption{Source parameters.\label{table:five}}
\begin{tabular}{lllll}
\hline\hline
\multicolumn{1}{c}{Parameter}           &
\multicolumn{1}{c}{MOA-2022-BLG-033}    &
\multicolumn{1}{c}{KMT-2023-BLG-0119}   &
\multicolumn{1}{c}{KMT-2023-BLG-1896}   \\
\hline
 $(V-I, I)$                 &  $(2.047 \pm 0.007, 18.970 \pm 0.004)$    &  $(1.890 \pm 0.037, 20.692 \pm 0.003)$   &  $(2.768 \pm 0.218, 23.485 \pm 0.017)$ \\
 $(V-I, I)_{\rm RGC}$       &  $(2.536, 16.918)                    $    &  $(1.980, 16.404)                    $   &  $(2.393, 15.807)                    $ \\
 $(V-I, I)_{{\rm RGC},0}$   &  $(1.060, 14.269)                    $    &  $(1.060, 14.502)                    $   &  $(1.060, 14.334)                    $ \\
 $(V-I, I)_0$               &  $(0.571 \pm 0.041, 16.321 \pm 0.020)$    &  $(0.970 \pm 0.055, 18.790 \pm 0.020)$   &  $(1.435 \pm 0.221, 22.011 \pm 0.026)$ \\
  Type                      &   F7V                                     &   K2.5V                                  &   K6V                                  \\
 $\theta_*$ ($\mu$as)       &  $1.473 \pm 0.119                    $    &  $0.738 \pm 0.066                    $   &  $0.246 \pm  0.057                   $ \\
 $\theta_{\rm E,min}$ (mas) &  0.25                                     &  0.07                                    &  0.25                                  \\
 $\mu_{\rm E,min}$ (mas/yr) &  0.79                                     &  0.44                                    &  1.65                                  \\
\hline
\end{tabular}
\end{table*}

The configurations for the lens system in KMT-2023-BLG-1896 are shown in Figure~\ref{fig:seven}. 
Although the anomaly has been determined to have a planetary origin, we also present the 
configuration for the binary solution to understand the cause of the similarity in the shape 
of the anomaly. The configurations for the planetary solutions are very similar to those of 
the previous events: for the inner solution, the source passed through the inner region between 
the central and planetary caustics, while for the outer solution, the source traversed the outer 
region of the caustics. The anomaly occurred when the source passed through the negative deviation 
region located at the rear side of the central caustic along the planet-host axis. In the binary 
model, the lens forms a Chang-Refsdal caustic \citep{Chang1979, Chang1984} with four folds 
converging at four cusps. The source passed through the region of reduced magnification between 
two cusps of the caustic, resulting in the negative deviation.

\section{Source stars} \label{sec:five}

In microlensing analysis, the main objective in characterizing the source is to determine the
angular Einstein radius. This is essential because the angular Einstein radius is linked to 
the physical lens parameters through the relation
\begin{equation}
\thetae = (\kappa M \pi_{\rm rel})^{1/2};\qquad \kappa = {4G \over c^2{\rm AU}}, 
\label{eq3}
\end{equation}
thereby offering an important constraint on the physical lens parameters. The angular Einstein
radius is determined from the normalized source radius through the relation
\begin{equation}
\thetae = {\theta_* \over \rho},
\label{eq4}
\end{equation}
where the angular radius of the source, $\theta_*$, can be inferred from the color and magnitude. 
In all analyzed events, finite-source effects were not detectable in the light curves, preventing 
the determination of $\rho$ and consequently the calculation of $\thetae$. Nevertheless, we define 
the source stars to ensure a comprehensive characterization of the events.

\begin{figure}[t]
\includegraphics[width=\columnwidth]{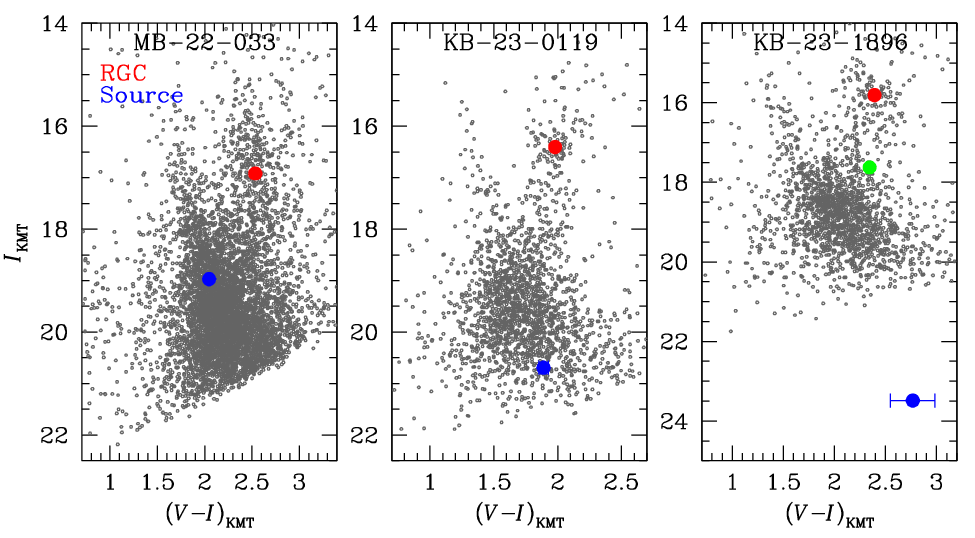}
\caption{
Locations of source stars (filled blue dots) in the instrumental color-magnitude diagrams. 
The red dot in each panel indicates the centroid of red giant clump (RGC).	
}
\label{fig:eight}
\end{figure}

We characterize the source by measuring its color ($V-I$) and magnitude ($I$). First, we 
determine the instrumental magnitudes in the $I$ and $V$ bands by regressing the photometric 
data processed using the pyDIA code \citep{Albrow2017} against the model. Next, we place the 
source in the instrumental color-magnitude diagram (CMD) for stars near the source. The color 
and magnitude are then calibrated using the centroid of the red giant clump (RGC), with their 
extinction and reddening-corrected 
values obtained from \citet{Bensby2013} for color and \citet{Nataf2013} for 
$I$-band magnitude.

Figure~\ref{fig:eight} shows the positions of the sources in the instrumental CMDs of the events, 
constructed from pyDIA photometry of stars in the KMTC image.  For KMT-2023-BLG-1896, for which 
the observed light curve was heavily influenced by flux from nearby blended stars, we also mark 
the position of the blend.  Table~\ref{table:five} lists the instrumental color and magnitude 
of the source, $(V-I, I)$, and those of the RGC centroid, $(V-I)_{\rm RGC}$. It also provides 
the de-reddened values for the RGC centroid, $(V-I)_{{\rm RGC},0}$, and the sources $(V-I)_0$. 
From these estimated colors and magnitudes, it is determined that the source is a late F-type 
star for MOA-2022-BLG-033, an early K-type star for KMT-2023-BLG-0119, and a mid K-type star 
for KMT-2023-BLG-1896. Although not utilized for $\thetae$ measurement, we also present the 
angular radii of the source stars. To estimate $\theta_*$, we first converted the $V-I$ color 
to $V-K$ using the \citet{Bessell1988} relation.  Subsequently, we derived $\theta_*$ based on 
the \citet{Kervella2004} relation between $(V-K, I)$ and $\theta_*$.

\section{Mass and distance to the planetary systems} \label{sec:six}

The physical parameters of a lens are determined through constraints provided by lensing 
observables.  Observables that can be measured from a lensing light curve include the event 
timescale ($\te$), the angular Einstein radius ($\thetae$), and the microlens parallax ($\pie$). 
The values of $\thetae$ and $\pie$ are related to the physical lens parameters by Eqs.~(\ref{eq3}) 
and (\ref{eq1}), respectively, and the event timescale is related by $\te= \thetae/\mu$. The 
basic observable of the event timescale was measured for all events, but the angular Einstein 
radius was not measured for any of them. For MOA-2022-BLG-033, the microlens parallax was 
additionally measured.

\begin{figure}[t]
\includegraphics[width=\columnwidth]{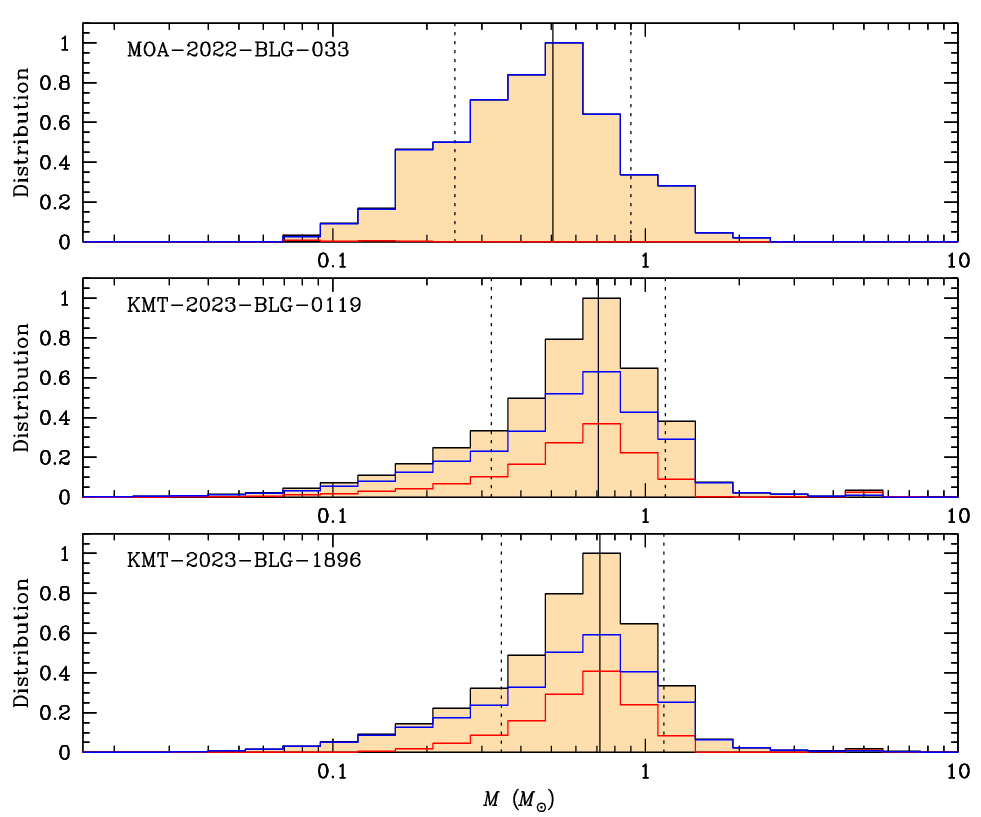}
\caption{
Posteriors of the mass of the planetary system. In each panel, the blue and red curves represent 
the distributions contributed by the disk and bulge lens populations, respectively, while the
black curve represents the combined distribution from both lens populations. The solid vertical line
denotes the mean of the distribution, and the dotted lines indicate the 1$\sigma$ uncertainty range.	
}
\label{fig:nine}
\end{figure}

\begin{table*}[t]
\caption{Physical lens parameters. \label{table:six}}
\begin{tabular}{lllll}
\hline\hline
\multicolumn{1}{c}{Parameter}           &
\multicolumn{1}{c}{MOA-2022-BLG-033}    &
\multicolumn{1}{c}{KMT-2023-BLG-0119}   &
\multicolumn{1}{c}{KMT-2023-BLG-1896}   \\
\hline
 $M_{\rm h}$ ($M_\odot$)    &  $0.51^{+0.39}_{-0.26} $           &  $0.71^{+0.45}_{-0.39}   $           &  $0.72^{+0.43}_{-0.37}  $          \\  [0.6ex]
 $M_{\rm p}$ ($M_{\rm E}$)  &  $12.15^{+9.41}_{-6.25}$           &  $83.70^{+53.46}_{-45.75}$           &  $16.35^{+9.81}_{-8.44}$          \\  [0.6ex]
 $\dl$ (kpc)                &  $1.84^{+1.00}_{-0.64} $           &  $5.41^{+2.29}_{-2.31}   $           &  $5.49^{+2.15}_{-2.30}  $          \\  [0.6ex]
 $a_{\perp}$ (AU)           &  $2.15^{+1.17}_{-0.75} $ (inner)   &  $3.14^{+1.32}_{-1.33}   $ (inner)   &  $2.86^{+1.12}_{-1.02}  $ (inner)  \\  [0.6ex]
                            &  $2.31^{+1.27}_{-0.81} $ (outer)   &  $4.01^{+1.69}_{-1.70}   $ (outer)   &  $4.62^{+1.81}_{-1.94}  $ (outer)  \\  [0.6ex]
 $P_{\rm disk}$             &   100\%                            &   68\%                               &   68\%                             \\  [0.6ex]
 $P_{\rm bulge}$            &   0\%                              &   32\%                               &   32\%                             \\
\hline
\end{tabular}
\end{table*}

We estimate the mass and distance to the planetary systems using a Bayesian analysis, incorporating 
the constraints provided by the measured observables of each event, along with the priors for the 
physical and dynamical distributions and the mass function of Galactic objects.  In this analysis, 
we first generated a large number of artificial lensing events through a Monte Carlo simulation. 
For each artificial event, the physical parameters $(M, \dl, \ds, )_i$ were derived from the priors 
of a mass function and and a Galaxy model.  We used the \citet{Jung2018} model for the mass function 
and the \citet{Jung2021} Galaxy model for the physical and dynamical distributions. Next, the 
observables $(\te, \thetae, \pie)_i$ corresponding to the physical parameters were computed. The 
posteriors of the physical lens parameters were then constructed by assigning a weight ($w_i$) to 
each event of
\begin{equation}
w_i = \exp \left( -{ \chi_i^2 \over 2}\right)\qquad 
\chi_i^2 = \chi_{t_{\rm E}, i}^2 + \chi_{\theta_{\rm E},i}^2  + \chi_{\pi_{\rm E},i}^2.
\label{eq5} 
\end{equation}
Here $\chi_{t_{\rm E}, i}^2 = (t_{{\rm E},i}-\te)^2/\sigma^2(\te)$, $\chi_{\theta_{\rm E}, i}^2 
= (\theta_{{\rm E},i}-\theta)^2/\sigma^2(\thetae)$, and $\chi_{\pi_{\rm E}, i}^2 = \sum_j \sum_k 
b_{j,k} (\pi_{{\rm E},j,i} - \pi_{{\rm E},j})  (\pi_{{\rm E},k,i} - \pi_{{\rm E},k})$, $[\te, 
\sigma(\te)]$ and $[\thetae, \sigma(\thetae)]$ represent the measured values of $\te$ and $\thetae$ 
and their associated uncertainties. The term $b_{j,k}$ denotes the inverse covariance matrix of 
$\pie$, $(\pi_{{\rm E},1}, \pi_{{\rm E},2})_i = (\pien, \piee)_i$, and $(\pien, \piee)$ indicates 
the measured microlens-parallax parameters.  Even though the angular Einstein radius values are not 
precisely measured for any event, we include a constraint on its minimum value, $\theta_{\rm E,min}$, 
in our analysis.

\begin{figure}[t]
\includegraphics[width=\columnwidth]{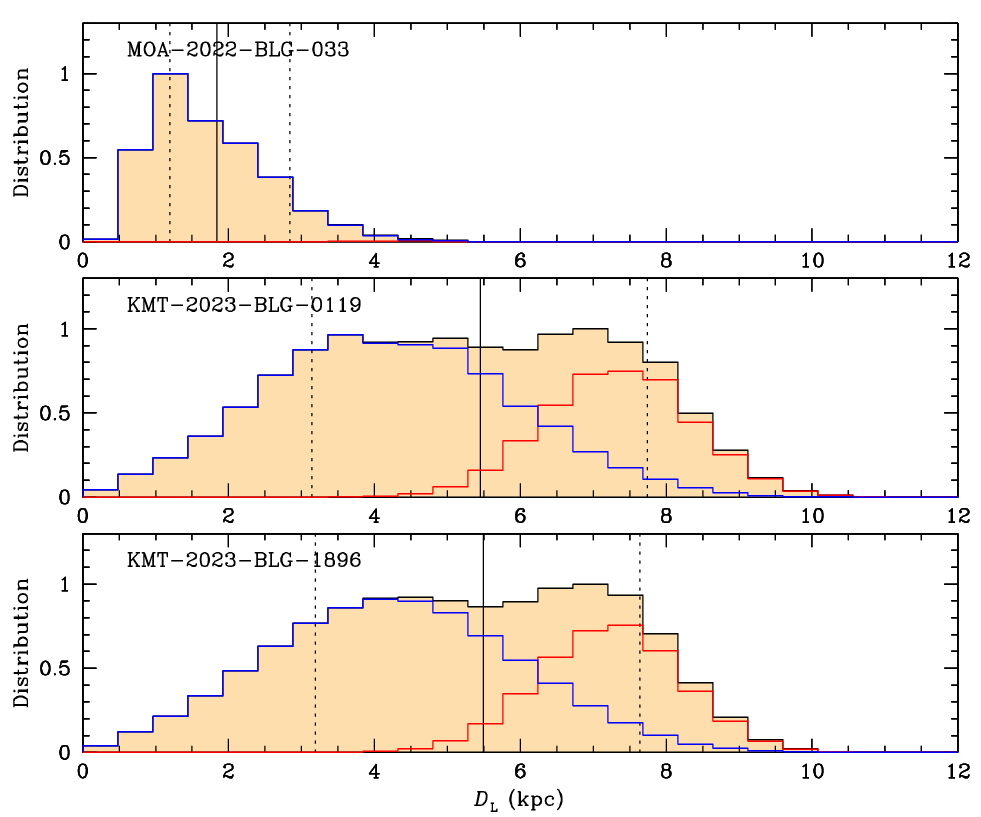}
\caption{
Posteriors of the distance to the planetary system. Notations are same as those in Fig.~\ref{fig:nine}.
}
\label{fig:ten}
\end{figure}

Figures~\ref{fig:nine} and \ref{fig:ten} display the constructed posteriors for the mass 
and distance of the planetary systems. The distance posterior range for MOA-2022-BLG-033 is 
significantly narrower than those for KMT-2023-BLG-0119 and KMT-2023-BLG-1896, due to the 
additional constraint provided by the measured microlens parallax. Blue and red curves 
illustrate the contributions of disk and bulge lenses, respectively. The lens of MOA-2022-BLG-033 
is highly likely to be located in the disk, whereas the contributions from both lens populations 
are roughly comparable for the other events.  The measured parallax of MOA-2022-BLG-033 strongly 
constrains the lens distance, but its constraint on the lens mass is relatively weak, though it 
still contributes to refining the mass estimate. Additionally, we found that the lower limit on 
$\thetae$ provides an insignificant constraint for any of the events.

Table~\ref{table:six} lists the estimated masses of the host ($M_{\rm h}$) and planet ($M_{\rm p}$), 
distance, and projected planet-host separation ($a_\perp$) for the planetary systems.  The lensing 
parameters for the pair of solutions resulting from the inner--outer degeneracy are similar to each 
other. Hence, we present the physical parameters estimated from the best-fit solution, except for 
the planet-host separation, which differs between the inner and outer solutions.  MOA-2022-BLG-033L 
is identified as a planetary system with an ice giant, approximately 12 times the mass of Earth, 
orbiting an early M dwarf star.  The companion of KMT-2023-BLG-1896L is also an ice giant, with a 
mass around 16 Earth masses, and orbits a mid-K-type main-sequence star.  The companion of 
KMT-2023-BLG-0119L, which has a mass about six times that of Uranus, orbits a mid-K-type dwarf star.  
The table also includes the probabilities of the planetary system being located in the disk 
($p_{\rm disk}$) or the bulge ($p_{\rm bulge}$).  The lens for MOA-2022-BLG-033 most likely situated 
in the disk. In contrast, for the other events, the chances of the lens being in the disk or the 
bulge are roughly equal.

\section{Summary and conclusion} \label{sec:seven}

We analyzed the anomalies in the light curves of the lensing events MOA-2022-BLG-033, 
KMT-2023-BLG-0119, and KMT-2023-BLG-1896. These anomalies share common traits, appearing near 
the peak of events with moderate to high magnification and displaying a distinctive short-term dip.

We conducted detailed modeling of the light curves to uncover the origin of the anomalies. 
This analysis revealed that all signals stem from planetary companions to the primary lens. 
The planet-to-host mass ratios are very low: approximately $q \sim 7.5 \times 10^{-5}$ for 
MOA-2022-BLG-033, $q \sim 3.6 \times 10^{-4}$ for KMT-2023-BLG-0119, and $q \sim 6.9 \times 
10^{-5}$ for KMT-2023-BLG-1896. The anomalies occurred as the source passed through the negative 
deviation region behind the central caustic along the planet-host axis. The solutions are subject 
to a common inner-outer degeneracy, resulting in slight variations in the estimated projected 
separation between the planet and its host.

We estimated the physical parameters of the planetary systems using Bayesian analyses based on 
the lensing observables. Although we measured the event timescale for all events, the angular 
Einstein radius was not determined for any. Additionally, the microlens parallax was only measured 
for MOA-2022-BLG-033.  Our analysis reveals that MOA-2022-BLG-033L hosts an ice giant, approximately 
12 times the mass of Earth, orbiting an early M dwarf star.  KMT-2023-BLG-1896L also features an 
ice giant, with a mass around 16 times that of Earth, orbiting a mid-K-type main-sequence star.  
The companion of KMT-2023-BLG-0119L, which is about the mass of Saturn orbits a mid-K-type dwarf 
star.  The lens for MOA-2022-BLG-033 is most likely located in the disk, whereas for the other 
events, the likelihood of the lens being in the disk or the bulge is roughly equal.

\begin{acknowledgements}
This research has made use of the KMTNet system operated by the Korea Astronomy and Space Science Institute 
(KASI) at three host sites of CTIO in Chile, SAAO in South Africa, and SSO in Australia. Data transfer from 
the host site to KASI was supported by the Korea Research Environment Open NETwork (KREONET). This research 
was supported by KASI under the R\&D program (project No. 2024-1-832-01) supervised by the Ministry of 
Science and ICT.
The MOA project is supported by JSPS KAKENHI Grant Number JP24253004, JP26247023, JP23340064, 
JP15H00781, JP16H06287, JP17H02871 and JP22H00153.
J.C.Y. and I.-G.S. acknowledge support from U.S. NSF Grant No. AST-2108414.
J.C.Y. acknowledges support from a Scholarly Studies grant from the Smithsonian Institution.
Y.S. acknowledges support from BSF Grant No. 2020740.
C.R. was supported by the Research fellowship of the Alexander von Humboldt Foundation.
W.Zang, and H.Y. acknowledge support by the National Natural Science Foundation of China (Grant No. 12133005). 
W.Zang acknowledges the support from the Harvard-Smithsonian Center for Astrophysics through the CfA 
Fellowship.  J.C.Y. and I.-G.S. acknowledge support from U.S. NSF Grant No. AST-2108414.
\end{acknowledgements}


\begin{thebibliography}{}
\bibitem[Alard \& Lupton(1998)]{Alard1998} Alard, C., \& Lupton, R. H. 1998, \apj, 503, 325	
\bibitem[Albrow et al.(2009)]{Albrow2009} Albrow, M., Horne, K., Bramich, D. M., et al. 2009, \mnras, 397, 2099
\bibitem[Albrow et al.(2017)]{Albrow2017} Albrow, M. 2017, MichaelDAlbrow/pyDIA: Initial Release on Github,Versionv1.0.0, Zenodo, doi:10.5281/zenodo.268049
\bibitem[Bensby et al.(2013)]{Bensby2013}  Bensby, T. Yee, J.C., Feltzing, S. et al. 2013, \aap, 549, A147
\bibitem[Bessell \& Brett(1988)]{Bessell1988} Bessell, M. S., \& Brett, J. M. 1988, \pasp, 100, 1134
\bibitem[Bond et al.(2001)]{Bond2001}  Bond, I. A., Abe, F., Dodd, R. J., et al. 2001, \mnras, 327, 868
\bibitem[Chang \& Refsdal(1979)]{Chang1979} Chang, K., \& Refsdal, S. 1979, Nature, 282, 561
\bibitem[Chang \& Refsdal(1984)]{Chang1984} Chang, K., \& Refsdal, S. 1984, \aap, 132, 168
\bibitem[Chung et al.(2005)]{Chung2005} Chung, S.-J., Han, C., Park, B.-G., et al. 2005, \apj, 630, 535	
\bibitem[Gaudi(1998)]{Gaudi1998} Gaudi, B. S. 1998, \apj, 506, 533
\bibitem[Gaudi \& Gould(1997)]{Gaudi1997} Gaudi, B. S., \& Gould, A. 1997, \apj, 486, 85
\bibitem[Gonzalez et al.(2012)]{Gonzalez2012} Gonzalez, O.~A., Rejkuba, M., Localize, M., et al.\ 2012, \aap, 543, A13
\bibitem[Gould(1992)]{Gould1992} Gould, A. 1992, \apj, 392, 442
\bibitem[Gould(2000)]{Gould2000} Gould, A. 2000, \apj, 542, 785
\bibitem[Gould(2004)]{Gould2004} Gould, A. 2004, \apj, 606, L319
\bibitem[Gould(2022)]{Gould2022a} Gould, A. 2022, arXiv:2209.12501
\bibitem[Gould \& Loeb(1992)]{GouldLoeb1992} Gould, A., \& Loeb, L. 1992, \apj, 396, 104	
\bibitem[Gould et al.(2022)]{Gould2022b} Gould, A., Han, C., Zang, W., et al. 2022, \aap, 664, A13
\bibitem[Griest \& Safizadeh(1998)]{Griest1998} Griest, K., \& Safizadeh, N. 1998, \apj, 500, 37
\bibitem[Han(2006)]{Han2006} Han, C. 2006, \apj, 638, 1080
\bibitem[Han et al.(2021b)]{Han2021a} Han, C., Udalski, A., Kim, D., et al. 2021a, \aap, 650, A89 
\bibitem[Han et al.(2021a)]{Han2021b} Han, C., Udalski, A., Kim, D., et al. 2021b, \aap, 655, A21 
\bibitem[Han et al.(2022)]{Han2022}  Han, C., Kim, D., Gould, A., et al. 2022, \aap, 664, A33 
\bibitem[Han et al.(2023a)]{Han2023a} Han, C., Lee, C.-U., Zang, W., et al. 2023a, \aap, 674, A90 
\bibitem[Han et al.(2023b)]{Han2023b} Han, C., Lee, C.-U., Bond, I. A., et al. 2023b, \aap, 676, A97 
\bibitem[Han et al.(2024a)]{Han2024a} Han, C., Jung, Y. K., Bond, I. A., et al. 2024a, \aap, 683, A115 
\bibitem[Han et al.(2024b)]{Han2024b} Han, C., Bond, I. A., Lee, C.-U., et al. 2024b, \aap, 687, A225 
\bibitem[Han et al.(2024c)]{Han2024c} Han, C., Albrow, M. D., Lee, C.-U. 2024c, \aap, 689, A209
\bibitem[Herrera-Martin et al.(2020)]{Herrera2020} Herrera-Martin, A., Albrow, A., Udalski, A., et al. 2020, \aj, 159, 134
\bibitem[Hwang et al.(2022)]{Hwang2022} Hwang, K.-H., Zang, W., Gould, A., et al. 2022, \aj, 163, 43
\bibitem[Jung et al.(2021)]{Jung2018} Jung, Y. K., Udalski, A., Gould, A., et al. 2018, \aj, 155, 219
\bibitem[Jung et al.(2021)]{Jung2021} Jung, Y. K., Han, C., Udalski, A., et al. 2021, \aj, 161, 293
\bibitem[Jung et al.(2023)]{Jung2023} Jung, Y. K., Zang, W., Wang, H., et al. 2023, \aj, 165, 226
\bibitem[Kervella et al.(2004)]{Kervella2004} Kervella, P., Th\'evenin, F., Di Folco, E., \& S\'egransan, D. 2004, \aap, 426, 29
\bibitem[Kim et al.(2016)]{Kim2016} Kim, S.-L., Lee, C.-U., Park, B.-G., et al. 2016, JKAS, 49, 37
\bibitem[Mao \& Paczy\'nski(1991)]{Mao1991}  Mao, S., \& Paczy\'nski, B. 1991, \apj, 374, 37
\bibitem[Nataf et al.(2013)]{Nataf2013}  Nataf, D. M., Gould, A., Fouqu\'e, P. et al. 2013, \apj, 769, 88
\bibitem[Smith et al.(2003)]{Smith2003} Smith, M. C., Mao, S., \& Paczy\'nski, B. 2003, \mnras, 339, 925
\bibitem[Skowron et al.(2011)]{Skowron2011} Skowron, J., Udalski, A., Gould, A., et al. 2011, \apj, 738, 87
\bibitem[Sumi et al.(2003)]{Sumi2003} Sumi, T., Abe, F., Bond, I. A., et al. 2003, \apj, 591, 204
\bibitem[Tomaney \& Crotts(1996)]{Tomaney1996} Tomaney, A. B., \& Crotts, A. P. S. 1996, \aj, 112, 2872
\bibitem[Yang et al.(2024)]{Yang2024} Yang, H., Yee, J. C., Hwang, K.-H., et al. 2024, \mnras, 528, 11
\bibitem[Yee et al.(2012)]{Yee2012} Yee, J. C., Shvartzvald, Y., Gal-Yam, A., et al.\ 2012, \apj, 755, 102
\bibitem[Yee et al.(2021)]{Yee2021} Yee, J. C., Zang, W., Udalski, A., et al. 2021, \aj, 162, 180
\bibitem[Zhang et al.(2022)]{Zhang2022} Zhang, K., Gaudi, B. S., Bloom, J. S. 2022, NatAs, 6, 782
\bibitem[Zhu et al.(2014)]{Zhu2014} Zhu, W., Penny, M., Mao, S., Gould, A., \& Gendron, R. 2014, \apj, 788, 73 
\end{thebibliography}
\end{document}